\documentstyle[12pt]{article}
\newcommand{\pr}{\paragraph{}}
\newcommand{\be}{\begin{equation}}
\newcommand{\ee}{\end{equation}}
\newcommand{\bea}{\begin{eqnarray}}
\def\lsim{\mathrel{\rlap {\raise.5ex\hbox{$ < $}}
{\lower.5ex\hbox{$\sim$}}}}

\newcommand{\nn}{\nonumber}
\newcommand{\eea}{\end{eqnarray}}
\newcommand{\nd}[1]{/\hspace{-0.6em} #1}
\newcommand{\nk}{\noindent}
\begin{document}

\begin{titlepage}

\begin{flushright}
CERN-TH.6755/92 \\
ACT-24/92 \\
CTP-TAMU-83/92
\end{flushright}
\begin{centering}
\vspace{.1in}
{\large {\bf $CPT$ Violation in String-Modified Quantum Mechanics
and the Neutral Kaon System }} \\
\vspace{.4in}
{\bf John Ellis}, {\bf N.E. Mavromatos} and
{\bf D.V. Nanopoulos}$^{\dagger}$ \\

\vspace{.05in}
Theory Division, CERN, CH-1211, Geneva 23, Switzerland. \\
\vspace{.05in}
\vspace{.1in}
{\bf Abstract} \\
\vspace{.05in}
\end{centering}
{\small We show that $CPT$ is in general violated
in a non-quantum-mechanical way
in the effective
low-energy theory derived from string theory, as a result of
apparent world-sheet charge non-conservation induced by
stringy monopoles corresponding to target-space black hole
configurations. This modification of quantum
mechanics does not violate energy conservation.
The magnitude of this effective
spontaneous violation of $CPT$ may not be be far from the
present experimental sensitivity in the
neutral kaon system. We demonstrate that our previously
proposed stringy modifications to the quantum-mechanical
description of the neutral kaon system violate $CPT$,
although in a different way from that assumed in
phenomenological analyses within conventional quantum
mechanics. We constrain the novel $CPT$-violating
parameters using available data on $K_L \rightarrow 2\pi$,
$K_S \rightarrow 3\pi ^0$ and semileptonic $K_{L,S}$
decay asymmetries. We demonstrate that these data
and an approximate treatment of interference effects
in $K \rightarrow 2\pi $ decays are consistent
with a
{\it non-vanishing}
amount of
$CPT$ violation at a level
accessible to a new round of experiments, and
further data and/or analysis are required
to exclude the extreme possibility that
they dominate over $CP$ violation.
Could non-quantum-field theoretical and
non-quantum-mechanical $CPT$ violation
usher in the long-awaited era of string phenomenology?}

\paragraph{}
\par
\vspace{0.6in}
\begin{flushleft}
CERN-TH.6755/92 \\
ACT-24/92 \\
CTP-TAMU-83/92 \\
December 1992 \\
\end{flushleft}
\vspace{0.3in}

\noindent $^{\dagger}$ {\it Permanent address}:
Center for Theoretical Physics, Dept. of Physics, \\
Texas A \& M University, College Station, TX 77843-4242, USA, \\
{\it and}
Astroparticle Physics Group,
Houston Advanced Research Center (HARC), \\
The Woodlands, TX 77381, USA\\

\end{titlepage}
\newpage
\section{Introduction}
\pr
The $CPT$ theorem is one of the deepest results of Quantum
Field Theory \cite{lud}. It
is a consequence of Lorentz invariance and locality, as
well as quantum mechanics. The best experimental test of $CPT$ invariance
so far has been in the neutral kaon system, where the equality of
particle and antiparticle masses has been confirmed to better than one
part in $10^{17}$ \footnote{ For comparison, the equality of proton and
antiproton masses has been checked to one part in $10^{10}$.}.
On the other
hand, $CP$
invariance, which is by no mean sacred in Quantum Field Theory,
appears to be violated in the neutral kaon system \cite{cp}
at a level higher than
any possible violation of $CPT$  invariance \cite{pdg}.
\pr
The possibility of a violation of $CPT$  invariance has been raised in
various theoretical contexts that go beyond conventional local Quantum
Field Theory. One important example is quantum gravity: it has been
argued that conventional quantum mechanics and the axioms of Quantum
Field Theory cannot be maintained once one considers
topologically non-trivial quantum fluctuations in the space-time
background \cite{hawkbek}. It
has been argued that a mixed-state description must
be used, because of the inevitable loss of information across event
horizons, even at the microscopic level. As a corollary of this
observation, it has been argued \cite{page}
that $CPT$
invariance must be abandoned in
quantum gravity, or at least re-expressed in a weakened
form \cite{lag}.
\pr
String theory is the only candidate we have for a consistent quantum
theory of gravity, and serves as a unique laboratory for studying and
quantifying these suggestions. We have argued in a recent series of
papers that whilst the usual axioms of Quantum Field Theory apply on the
string world-sheet, and quantum coherence is maintained by the complete
spectrum of massive string states \cite{emn1}, the effective truncated
theory of light
particles observed in laboratory experiments must obey a modified form
of quantum mechanics which allows for the evolution of pure states into
mixed states \cite{emnqm}. With
this new string motivation, we have also revived \cite{emnkaons}
a
non-quantum-mechanical density matrix description of the neutral kaon
system that two of us (J.E. and D.V.N.) proposed several
years ago \cite{EHNS}
together with J.S. Hagelin and M. Srednicki in the general
context of quantum
gravity, and confronted it with some of the available experimental data
on the neutral kaon system \cite{pdg}.
\pr
We pointed out in this recent paper some possible tests of this
non-quantum-mechanical density matrix
formalism in the neutral kaon system, including for example
comparisons between the experimental rate for
${K_L \rightarrow 2 \pi}$ decay
and semileptonic decay asymmetries for $K_L$ and $K_S$. As we also
pointed out, similar comparisons had been discussed previously as
probes of a possible violation of
$CPT$ \cite{peccei}. However, those previous
proposals
were in the context of a conventional quantum-mechanical pure
state vector
description of the neutral kaon system, and the correspondence with the
parameters of our modified mixed-state
density matrix description was neither
direct nor obvious.
\pr
The possibility that $CPT$ might be violated in string theory has been
discussed by several authors. It has been pointed out that $CPT$
invariance is linked to the conservation of charges on the string
world-sheet, and it has been shown that these charges are indeed
conserved in a flat space-time background \cite{wittcpt,sonoda}. It
has been conjectured that
$CPT$ might be spontaneously violated in string
field theory \cite{kost}, but
charge conservation on the world-sheet and $CPT$ invariance in
space-time have not been studied in the type of non-trivial space-time
background which, according to the quantum gravity literature, might
lead to $CPT$ violation.
\pr
We have developed in previous papers
a theoretical framework suitable for
addressing the problem of $CPT$ violation in topologically
non-trivial space-time backgrounds in string theory. It has been shown
that physics in the neighbourhood of a spherically-symmetric stringy
black hole singularity \cite{witt}
in four-dimensional space-time can be represented
by an Abelian $U(1)$ Chern-Simons gauge theory with a monopole
configuration on the world-sheet \cite{emndua}. The
time evolution of a black hole is
described by a homotopic extension of the $1+1$-dimensional gauge
theory to $2+1$ dimensions \cite{emnhall}. The
quantum foam of microscopic topologically
non-trivial space-time fluctuations consists of a gas of
monopole-antimonopole pairs. Furthermore, the infinite symmetries
coupling observable light particles to unobserved massive solitonic
string states associated with these world-sheet
monopoles implied \cite{emnqm}
that the observable particles should be described as an open quantum
system described by the modified
quantum-mechanical formalism \cite{EHNS}
mentioned above.
\pr
In this paper we use this framework to show that $CPT$ violation
is generic in the effective low-energy theory derived from string
theory, once topologically non-trivial space-time fluctuations are
taken into account. We show that the above-mentioned Chern-Simons
monopoles violate charge conservation on the world-sheet, leading to
non-quantum-field-theoretical
effects
in the truncated low-energy
space-time theory that violate $CPT$ but conserve energy.
We attempt to
quantify the possible magnitude of such effects, and confirm our
previous estimate \cite{emnqm}
that they could be suppressed by just one power of
the light particle mass scale divided by the Planck mass, i.e.,
possibly of order $10^{-19}$ in the dynamics of strongly-interacting
particles. Since this estimate is very close to the present
experimental upper limits on $CPT$
violation in the neutral kaon system \cite{pdg},
we re-examine these previous analyses in the context of our density
matrix formalism for effectively-open quantum systems. We demonstrate
the existence of three extra $CPT$-violating parameters distinct from
particle-antiparticle mass and lifetime differences that appear in the
conventional state vector description of the neutral kaon system. We
then make a preliminary analysis of various aspects of the available
data within this modified quantum framework, including
measurements of the
$K_L \rightarrow 2 \pi$ decay rate, published
(preliminary \cite{CPLEAR}) data on $K_L$
($K_S$) semileptonic decay asymmetries,
the experimental upper
bound on $K_S \rightarrow 3 {\pi}^0$ decays, and
an estimate of the likely
sensitivity of intermediate-time
interference measurements \cite{NA31}
of the
phases of $K \rightarrow 2 \pi$ amplitudes. A full analysis of the latter
constraint would require a global fit to all the available experimental
data within the density matrix formalism, which would take us beyond
the scope of this paper.
\pr
Our analysis does not exclude the
possibility that
the apparent $CP$ violation observed
in
$K_L \rightarrow 2 \pi$ decays and semileptonic decay
asymmetries
could be accompanied (or even replaced) by
non-quantum-mechanical
$CPT$-violating
effects. This radical possibility could be constrained
by a complete fit
to the intermediate-time data, or by
improvements in the present
measurements of the $K_S$ semileptonic decay asymmetry, or in the
present bounds on the $K_S \rightarrow 3 {\pi^0}$ decay rate. The
latter could
be achieved by the CPLEAR experiment at CERN or the forthcoming
DA$\Phi$NE
project \cite{dafnehb} at Frascati.
\pr
\section{Spontaneous $CPT$ Violation and Modifications of Quantum
Mechanics in String Theory}
\pr
We are concerned in this paper with the truncated effective theory of the
light degrees of freedom in a generic string theory, and in particular
with their quantum time evolution. These light string degrees of freedom
are linked to the massive string modes by an infinite set of gauge
symmetries that mix mass-levels \cite{venkub}. Among these, we have
identified \cite{emn1}
a
$W_{\infty}$ symmetry that is responsible for the maintenance of quantum
coherence in the presence of stringy black holes \cite{witt},
precisely as a result of
this coupling of different mass-levels. Since the effective low-energy
theory is a truncation of the full string theory, it is in general
subject to non-trivial renormalization effects. For example,
shifts in the ``tachyon", i.e. light particle, background are not exactly
marginal deformations of the conformal Wess-Zumino coset theory that
describes a spherically-symmetric
black hole in space-time \cite{chaudh}. This means
that the couplings $g^i$ of the tachyons have non-trivial renormalization
group coefficients ${\beta}^i$. We have proposed that target time $t$ be
identified with the corresponding world-sheet renormalization scale.
\pr
A conventional laboratory experiment, measuring for example neutral kaon
decays, does not detect massive string states, although these could in
principle be observed by an infinite set of generalized Aharonov-Bohm
phase measurements \cite{emnab}. In
the absence of such a measurement, any
observation in the effective low-energy theory will appear as a
non-conformal deformation, which will lead to a renormalization group
flow of the truncated light-mode system. Unitarity of the effective
light-particle theory implies that this flow is irreversible, and the
identification of the flow variable with $t$ provides an arrow of time,
i.e. spontaneous T-violation, in the string
universe \cite{emnhall}. {\it
Time is a
statistical parameter that measures the interaction (gravitational
friction) of the light particles with the massive string modes in the
presence of singular space-time backgrounds (foam)}.
\pr
The associated quantum time evolution of the light-mode density matrix is
given by a modified Liouville equation \cite{EHNS,emnqm}:
\be
  \dot \rho = i[\rho, H] +i G_{ij} [\rho , q^i]\beta ^j
\label{qmv}
\ee

\nk where $H$ is
the light-mode Hamiltonian and $G_{ij}$ is the Zamolodchikov
metric in the coupling space \{$g^i$ \}.
Associated with the evolution (1) is a
monotonic increase in entropy
\be
 \dot S =   Tr \beta^i G_{ij} \beta ^j
\frac{\partial \rho}{\partial H} ln\rho
\label{entrop}
\ee

\nk It has
been observed \cite{EHNS,gross}
that such a modification (1) of the Liouville equation
will
in general lead to the non-conservation of quantities associated with
symmetries, and this has been held against \cite{banks}
such a modification of
conventional
quantum mechanics. {\it We take this opportunity to point out that
energy
is indeed conserved by the modification (1), for specifically
stringy reasons}. It is easy
to derive from (1) the following expresion for
the time-variation of the expectation value of the light-system
Hamiltonian $H$:
\be
 \partial _t <<H>>
 = Tr({\dot \rho}H )+Tr(\partial _t H \rho) =Tr(
  i[g^i, H] \beta^jG_{ij}\rho ) + \dot H \rho)
\label{energ}
\ee

\nk where $<< O >> \equiv Tr (O \rho) $ denotes the
average
value of the observable $O$
in this non-quantum mechanical
framework.
Identifying -$i [g^i,H]$ with the time derivative ${\beta}^i$ of the
coupling $g^i$, equation (3) becomes
\be
 \partial _t <<H>> = Tr({\dot H}\rho -\beta^iG_{ij}\beta^j\rho)=
<< \partial _t (H+ C)>>
\label{zamol}
\ee

\nk where
$C$ is the Zamolodchikov C-function \cite{zam}. Since this can
be identified with the string effective action \cite{mavmir},
$H$ + $C$ must be a
constant, since $\partial _t (C+H) \propto \partial _t (p_i\beta^i)=0$,
as a result of the fact that
neither $p_i$ nor
${\beta}^i$ has any explicit cutoff dependence, because the world-sheet
theory is finite.
\pr
However, space-time foam does lead to the apparent violation of certain
global symmetries
on the world-sheet, which leads in turn to $CPT$ violation
in the effective low-energy theory. This can be seen using the Hall fluid
picture of space-time foam discussed in ref. \cite{emnhall}.
According to this
picture, space-time foam can be represented as a statistical population
of topological defects (spikes) on the world-sheet, which correspond to
monopoles of a $2+1$-dimensional Abelian $U(1)$ Chern-Simons gauge theory
introduced as a homotopic extension of the underlying $1+1$-dimensional
world-sheet theory. The effective $2+1$-dimensional action is
\be
  \int _{\Sigma \otimes S^1} [|D(a)\phi |^2   + \frac{k}{4\pi}
  \varepsilon _{\mu\nu\rho} a_\mu\partial _\nu a_\rho+ V(\phi)]
\label{cshiggs}
\ee

\nk
where $a_{\mu}$ is the
Chern-Simons gauge field coupled to a complex
scalar field $\phi$ that represents deviations from the singularity, and
hence the generation of space-time through symmetry breaking provided by
a suitable form of the effective potential $V(\phi)$. The parameter $k$
is the Wess-Zumino level parameter, and the string black hole corresponds
to the adiabatic limit in which the pseudo-``temperature'' $\tau$ =
${\beta}^{-1}$, where $\beta$ is the radius of the compactified $S^1$ in
(5), goes to zero. Non-critical deformations correspond to deviations of
$k$ from its critical value $9/4$. The black hole is a non-topological
soliton in this picture, which can be regarded as a monopole-instanton in
the $2+1$-dimensional
effective theory (5). It is known \cite{CSmon}
that the charge of
the monopole is not in
general conserved in $2+1$-dimensional Chern-Simons
theory, in the sense that there are tunnelling processes between states
with different monopole charges. Rather like the (more?) familiar
instantons in $4$-dimensional gauge theory \cite{thooft}, the $a_{\mu}$
monopole-instantons lead to charge-violating effective interactions that
look like
\be
    <f|e^{-HT}|i> \propto  \int e^{-S_{eff}}  \qquad ; \qquad
S_{eff} =\int K e^{-B _i+\alpha} \Phi _{em}^q + h.c
\label{eff}
\ee

\nk in the
low-energy limit. Here ${\Phi}_{em}$ is a gauge-invariant charge-
and magnetic-flux-changing operator in the Chern-Simons theory, $q$
denotes the monopole charge, $B$ is the classical instanton action
suppression
term, $K$ is the one-loop quantum correction, and $\alpha$ is
an arbitrary phase.
\pr
The charge violation due to this type of tunnelling process is similar to
that expected from wormholes \cite{coleman}
in $4$-dimensional Eulidean gravity. In our
case, a tunnelling event between states with monopoles of different
masses, which are the same as the monopole charges $q$ since the
monopoles are always extremal, leads
locally to apparent charge violation by
the space-time foam. In
the wormhole case, it is possible to transfer charge
through the wormhole from
one region of space-time to another, also with a
local , but not global, violation of charge. In our case, the charge is
transferred to to the black hole, i.e. to massive string modes according
to the selection rules discussed in ref. \cite{emnab}.
\pr
The apparent charge violation found above
manifests
itself as a violation
of $CPT$ in the effective low-energy theory, as follows
from the general discussion in
ref. \cite{wittcpt}. From the
point of view of the world-sheet, this
is spontaneous $CPT$ violation. It
appears $only$ associated with the
non-quantum-mechanical open-system term
in the quantum evolution equation (1), and $not$ within the context of
conventional quantum mechanics and Quantum Field Theory, as assumed in
previous phenomenological analyses \cite{peccei}.
It is allowed because the conditions
used to prove the $CPT$ theorem in Quantum Field Theory
are not all met in
string theory: specifically, string theory is non-local. This deviation
from locality is sufficiently weak to preserve $CPT$ in flat target
space-times \cite{wittcpt,kost}. However,
the non-locality rises up to violate $CPT$ in a black
hole background, and hence once space-time foam is taken into account, as
has been argued on general grounds in the
context of quantum gravity \cite{page}.
\pr
The next problem is to estimate the possible order of magnitude of
$CPT$-violating effects. The non-quantum-mechanical open-system term in
equation (1)
contains an explicit coordinate factor $q^i$, which one would
naively expect to be of order ${m_P}^{-1}$, corresponding to the scale of
fluctuations in the space-time foam. Are there any other Planckian
suppression factors? This inverse linear dependence on $m_P$ mirrors the
inverse linear
dependence on the scales of the ``environmental'' oscillators
in the Feynman-Vernon \cite{fv,cald}
formulation of open systems, to which our formalism
is very similar \cite{emnqm}. In
our case, the key issue is the scale size of the
microscopic black hole fluctuations in the space-time foam, which is in
turn related to the dominant range of values of the pseudo-``temperature"
${\beta}^{-1}$. We recall that the world-sheet has both spikes and
dual vortices, which are viewed \cite{thomas} as statistical
excitations of
the system characterized by an extra scale,
the pseudo-``temperature'' $\beta ^{-1}$, which does not
have a literal thermal interpretation, but
parametrizes
the
phase diagram for the
two-dimensional gas of topological defects.
Minkowski black holes correspond to
pseudo-``temperatures" below the
Berezinskii\cite{vlb}-Kosterlitz-Thouless \cite{kt} (BKT)
temperature above which
the monopoles become free. The conformal dimensions of the
operators that describe deformations due to monopoles are
\be
     4\Delta _m = \frac{e^2}{2\pi \beta}
\label{monop}
\ee

\nk  where
the critical
temperature corresponding to the BKT transition is that
associated with ${\Delta}_m$ = $1$.
For $e=1$, which is energetically preferred
this implies a critical temperature
$T_c =     8\pi M_P$. The charge quantization
condition for a monopole with charge $q_m$
\be
2q_m \pi\beta =e=1
\label{quantiz}
\ee

\nk then allows for black hole configurations with masses
up to $\simeq 8\sqrt{2}M_P$ in the space-time foam,
since
$M_{bh} \propto 2\sqrt{2}q_m $.  Tunneling effects then
restrict the
physics of space-time foam
to the region of black-hole masses
$1 \le M_{bh}/M_P \le 8\sqrt{2} $.
The correlation
functions that appear implicitly in (1) via the Zamolodchikov
metric $G_{ij}$ and the renormalization group coefficients
contain logarithmic dependences on the black hole mass that
could vanish for some particular $M_{BH} =O(M_P)$, but will
be $O(1)$ in general. We have not yet identified
any exponential or power Planckian suppression beyond
the single factor of $M_P^{-1}$ already mentioned.
This means that $CPT$ violation is generically {\it maximal} :
non-quantum mechanical effects should be able to violate
$CPT$ as much as the weak interactions violate $C$ and $P$.
\pr
However,
we cannot rule out the possibility that there might be
extra suppressions, at least for monopole matrix elements
$(6)$
between
specific final states, in the same way as 4-dimensional
Euclidean instanton effects are suppressed by extra factors
of light fermion masses $m_f$ if the external states do not
correspond precisely to zero modes of the instanton \cite{thooft}.
Specifically, there could be some selection rule, which
we cannot yet identify, that suppresses $CPT$-violating
effects in neutral kaons. However, we find it sufficiently
interesting that we have identified a possible process
in string theory whereby $CPT$ and $T$ could be violated
spontaneously at similar rates, and possibly
within reach of present and future kaon experiments,
that we now go on to explore further
the phenomenology \cite{emnkaons}
of our
non-quantum-mechanical mechanism for $CPT$-violation.

\pr
\section{$CPT$ Violation in the Quantum-Mechanical Density Matrix
Formalism for Neutral Kaons}
\pr
Fortified by the above argument that $CPT$ should indeed be violated in
the effective low-energy theory derived from string, and the
possibility that $CPT$ violation parameters might not be much smaller
than the present experimental limits from the neutral kaon system, we
now review the density matrix formalism \cite{EHNS}
for neutral kaons, and
analyze the possibility of $CPT$ violation, initially within the
conventional quantum-mechanical framework. The time evolution of a
generic density matrix is determined by the equation
\be
\partial _t \rho = -i (H\rho - \rho H^{\dagger}) + \nd{\delta H} \rho
\label{denmatr}
\ee

\nk where the open-system $\nd{\delta H}$ term is absent in conventional
quantum mechanics. The conventional phenomenological Hamiltonian for the
neutral kaon system contains hermitian (mass) and antihermitian (decay)
components:
\be
  H = \left( \begin{array}{c}
 (M + \frac{1}{2}\delta M) - \frac{1}{2}i(\Gamma + \frac{1}{2}
 \delta \Gamma)
   \qquad  \qquad
   M_{12}^{*} - \frac{1}{2}i\Gamma _{12} \\
           M_{12}  - \frac{1}{2}i\Gamma _{12}
    \qquad  \qquad
    (M - \frac{1}{2}
    \delta M)-\frac{1}{2}i(\Gamma
    - \frac{1}{2}
    \delta \Gamma ) \end{array}\right)
\label{hmatr}
\ee

\nk in the ($K^0$, ${\overline K}^0$) basis.
\pr
The ${\delta}M$ and ${\delta}{\Gamma}$ terms
violate $CPT$ \cite{peccei}. As in ref. \cite{EHNS},
we define components of $\rho$ and $H$ by
\be
\rho \equiv  \frac{1}{2}\rho _{\alpha} \sigma _{\alpha}
\qquad ; \qquad H \equiv \frac{1}{2}h_{\alpha}\sigma _{\alpha}
\label{rhosigma}
\ee

\nk in
a Pauli $\sigma$-matrix representation : the ${\rho_{\alpha}}$ are
real, but the $h_{\beta}$ are complex. The $CPT$ transformation is
represented by
\be
    CPT  |K^{0} >=e^{i\phi}|{\overline K}^{0}>, \qquad
 CPT  |{\overline K}^{0}>=e^{-i\phi}|K^{0}>
\label{cpt}
\ee

\nk for
some phase ${\phi}$, which is represented in our matrix formalism by
\be
   CPT  \equiv \left( \begin{array}{c}
 0   \qquad  e^{i\phi} \\
 e^{-i\phi}  \qquad 0 \end{array}\right)
\label{cptmatr}
\ee

\nk Since this matrix is a linear combination of ${\sigma}_{1,2}$, $CPT$
invariance of the phenomenological Hamiltonian, $H$ =
$(CPT)^{-1}  H  CPT$, clearly requires that $H$ contain no term
proportional to ${\sigma}_3$, i.e., $h_3$ = $0$ so that ${\delta}M$ =
${\delta}{\Gamma}$ = $0$.
\pr
Conventional quantum-mechanical evolution is represented by
$\partial _t \rho =
h_{{\alpha}{\beta}}{\rho_{\beta}}$, where, in the ($K^0$,
${\overline K}^0$)
basis and allowing for the possibility of $CPT$ violation,
\be
  h_{\alpha\beta} \equiv \left( \begin{array}{c}
 Imh_0 \qquad Imh_1 \qquad Imh_2 \qquad Imh_3 \\
 Imh_1 \qquad Imh_0 \qquad -Reh_3 \qquad Reh_2 \\
 Imh_2 \qquad Reh_3 \qquad Imh_0 \qquad -Reh_1 \\
 Imh_3 \qquad -Reh_2 \qquad Reh_1 \qquad Imh_0 \end{array}\right)
\label{habmatr}
\ee

\nk Now is an appropriate time to transform to the $K_{1,2}  =
\frac{1}{\sqrt{2}}(K^0 \mp {\overline K}^0)$
basis, corresponding to $ \sigma _1  \leftarrow\rightarrow
  \sigma _3$, $\sigma_2 \leftarrow\rightarrow -\sigma _2 $,
  in which $h_{\alpha\beta }$ becomes

\be
 h_{\alpha\beta}
 =\left( \begin{array}{c}  - \Gamma \qquad -\frac{1}{2}\delta \Gamma
\qquad -Im \Gamma _{12} \qquad -Re\Gamma _{12} \\
 - \frac{1}{2}\delta \Gamma
  \qquad -\Gamma \qquad - 2ReM_{12}\qquad  -2Im M_{12} \\
 - Im \Gamma_{12} \qquad  2ReM_{12} \qquad -\Gamma \qquad -\delta M    \\
 -Re\Gamma _{12} \qquad -2Im M_{12} \qquad \delta M   \qquad -\Gamma
\end{array}\right)
\label{hcomp}
\ee

\nk The corresponding equations of motion for the components of $\rho$ in
the $K_{1,2}$ basis are
\bea
\nonumber  \partial _t
\rho_{11} &=& -(\Gamma + Re \Gamma _{12} )\rho _{11}
-(2ImM_{12} + \frac{1}{2}
\delta \Gamma )Re \rho _{12}
-(Im \Gamma _{12} -  \delta M) Im \rho _{12} \\
\nonumber \partial _t \rho _{12} &=&
-(\Gamma - 2i Re M_{12} )\rho _{12}
+ (Im M_{12} - \frac{1}{2}i Im \Gamma _{12} - \frac{1}{2}
\delta \Gamma -i\delta M)\rho _{11} \\
\nonumber &-& (Im M_{12}
 +  \frac{1}{2}i Im \Gamma _{12})\rho _{22} \\
\nonumber \partial _t \rho _{22}
&=& -(\Gamma - Re \Gamma _{12} )\rho _{22}
+ (2 Im M_{12} - \frac{1}{2}
\delta \Gamma )Re \rho _{12} \\
&-& (Im \Gamma _{12} +  \delta M)Im \rho _{12}
\label{rhoeq}
\eea

\nk It is easy to check that ${\rho}$ decays at large $t$ to
\be
  \rho \simeq e^{-\Gamma _Lt}
 \left( \begin{array}{c}
 1   \qquad  \epsilon^* + \delta^* \\
 \epsilon + \delta \qquad |\epsilon + \delta |^2 \end{array}\right)
\label{rhodec}
\ee

\nk corresponding to a pure long-lived mass eigenstate $K_L$, with the
CP-violating parameter ${\epsilon}$ given by \cite{dafnehb}
\be
  \epsilon =\frac{\frac{1}{2}i Im \Gamma _{12} - Im M_{12}}
{\frac{1}{2} \Delta \Gamma - i\Delta M }
\label{epsilon}
\ee

\nk
where $ \Delta M$ = $M_L$ - $M_S$ is positive and $ \Delta  \Gamma   =
  \Gamma _L -   \Gamma _S$ is negative, and the $CPT$-violating parameter
${\delta}$ by
\be
\delta    \simeq -\frac{1}{2}
\frac{ \frac{1}{2}\delta \Gamma - i\delta M}
{\frac{1}{2} \Delta \Gamma - i\Delta M}
\label{delta}
\ee

\nk Conversely, in the short-$t$ limit  a $K_S$ state is represented
by
\be
  \rho \simeq e^{-\Gamma _St}
 \left( \begin{array}{c}
 |\epsilon - \delta |^2    \qquad  \epsilon^* - \delta^* \\
 \epsilon - \delta \qquad 1  \end{array}\right)
\label{rhodecs}
\ee

\nk where
we see that the relative
signs of the ${\delta}$ terms have reversed: this
is the signature of $CPT$ violation in the conventional
quantum-mechanical formalism. Note that
the density matrices
(17, 20) correspond to the state vectors
\be
|K_{L(S)}> \propto ((1 + \epsilon \mp  \delta )|K^0> \mp
 (1 - \epsilon \pm  \delta )|{\overline K}^0>)
\label{kaons}
\ee

\nk and are both pure,
as should
be expected in conventional quantum mechanics, even if $CPT$ is
violated.
\pr
\section{$CPT$ Violation in the String Modification of the Density Matrix
Formalism}
\pr
We now extend the above formalism to include the
non-quantum-mechanical term $\nd{\delta H}$ in equation (1). This can be
parametrized by a $4\times 4$ matrix
$\nd{h}_{\alpha\beta} $ analogous to
the matrix $h_{\alpha\beta }$ discussed above. We work in the
$K_{1,2}$ basis. As discussed in ref. \cite{EHNS}, we
assume that the dominant
violations of quantum mechanics conserve strangeness, so that
$\nd{h}_{1\beta }$ = 0, and hence that $\nd{h}_{0\beta }$ = 0 so as to
conserve probability. Since $\nd{h}_{\alpha\beta }$ is a symmetric
matrix, it follows that also $\nd{h}_{\alpha 0}$ = $\nd{h}_{\alpha 1}$
= 0. Moreover, $\nd{h}_{\alpha\beta }$ must be a negative matrix, so
we arrive at the general parametrization
\pr
\be
  {\nd h}_{\alpha\beta} =\left( \begin{array}{c}
 0  \qquad  0 \qquad 0 \qquad 0 \\
 0  \qquad  0 \qquad 0 \qquad 0 \\
 0  \qquad  0 \qquad -2\alpha  \qquad -2\beta \\
 0  \qquad  0 \qquad -2\beta \qquad -2\gamma \end{array}\right)
\label{nine}
\ee

\nk where $\alpha$, $\gamma  > 0$, ${\alpha}{\gamma}  >
{\beta}^2$ \cite{EHNS}.
\pr
We recall that the $CPT$ transformation, which is [see (\ref{cptmatr})]
a linear
combination of $ \sigma _{1,2}$ in the ($K^0$, ${\overline K}^0$)
basis, becomes
in the $K_{1,2}$ basis a linear combination of $ \sigma _{3,2}$. It is
apparent that none of the non-zero terms $\propto   \alpha ,  \beta ,
 \gamma $ in $\nd{h}_{\alpha\beta}$
commutes with the $CPT$ transformation. In other words, each of the three
parameters $\alpha$, $\beta$, $\gamma$ violates $CPT$, leading to a much
richer phenomenology than in conventional quantum mechanics. This is
because the symmetric $\nd{h}$ matrix has three parameters in its
bottom right-hand $2\times 2$ submatrix, whereas
the antisymmetric $h$ matrix
has only one complex $CPT$-violating parameter. This means that the
experimental constraints \cite{pdg}
on $CPT$ violation have to be rethought, as we
discuss in section 5.
\pr
The equations of motion for the
components of $\rho$ in the $K_{1,2}$
basis are \cite{EHNS,emnkaons}
\bea
\nonumber  \partial _t
\rho_{11} &=& -(\Gamma + Re \Gamma _{12} )\rho _{11}
-\gamma (\rho _{11}-\rho _{22} ) - 2Im M_{12} Re \rho _{12}
-(Im \Gamma _{12} + 2\beta ) Im \rho _{12} \\
\nonumber \partial _t \rho _{12} &=&
-(\Gamma - 2i Re M_{12} )\rho _{12} - 2i \alpha Im \rho _{12}
+ (Im M_{12} - \frac{1}{2}i Im \Gamma _{12} - i\beta )\rho _{11} \\
\nonumber &-& (Im M_{12}
 +  \frac{1}{2}i Im \Gamma _{12} - i\beta )\rho _{22} \\
\nonumber \partial _t \rho _{22}
&=& -(\Gamma - Re \Gamma _{12} )\rho _{22}
+ \gamma (\rho _{11} - \rho _{22} ) + 2 Im M_{12} Re \rho _{12} \\
&-& (Im \Gamma _{12} - 2\beta )Im \rho _{12}
\label{rhoeq2}
\eea

\nk
which are to be compared with the corresponding quantum-mechanical
equations (\ref{rhoeq}). We
see that the parameters ${\delta}M$ and $\beta$ play
similar roles, although they appear with different relative signs in
some places, because of the symmetry of $\nd{h}$ as opposed to the
antisymmetry of $h$.
\pr
These differences are important for the asymptotic limits of the
density matrix, and its impurity. It is easy to check that, for large
$t$, $\rho$ decays exponentially to
\pr
\be
\rho _L
\propto \left( \begin{array}{c} 1 \qquad  \qquad
\frac{-\frac{1}{2}i  (Im \Gamma _{12} + 2\beta )- Im M_{12} }
{\frac{1}{2} \Delta \Gamma + i \Delta M } \\
\frac{\frac{1}{2}i (Im \Gamma _{12} + 2\beta )- Im M_{12} }
{\frac{1}{2}\Delta \Gamma  - i \Delta M} \qquad \qquad
|\epsilon |^2 + \frac{\gamma}{\Delta \Gamma } -
\frac{4\beta Im M_{12} (\Delta M / \Delta \Gamma ) + \beta ^2 }
{\frac{1}{4} \Delta \Gamma ^2 + \Delta M^2 } \end{array} \right)
\label{rhdeqm}
\ee

\nk where
the $CP$ impurity parameter $\epsilon$ is given by equation
(\ref{epsilon}) as
usual. The density matrix (\ref{rhdeqm}) describes
a mixed state corresponding to
a mixture of
conventional $K_L$ and
$K_S$ states, and not a pure
state as
in equation (17). Conversely, if we look in the short-time limit
for a solution of the
equations (\ref{rhoeq2}) with
${\rho}_{11}  <<  {\rho}_{12}  <<  {\rho}_{22}$,
we again find a
mixed state:
\be
 \rho _S
 \propto \left( \begin{array}{c} |\epsilon |^2 +
\frac{\gamma }{|\Delta \Gamma |} -
\frac{-4\beta Im M_{12} (\Delta M/\Delta \Gamma )+ \beta ^2 }
{ \frac{1}{4} \Delta \Gamma ^2 + \Delta M^2 } \qquad
\epsilon - \frac{i\beta}{\frac{\Delta \Gamma} {2} - i \Delta M} \\
\epsilon ^* + \frac{i\beta} {\frac{\Delta \Gamma}{2} +
i \Delta M} \qquad 1 \end{array} \right)
\label{thirteen}
\ee

\nk to be contrasted with the conventional pure $K_S$ state
(20).
\pr
\section{Phenomenological Analysis of Possible $CPT$ and Apparent $CP$
Violation in the Neutral Kaon System}
\pr
The framework for treating experimental observables in our
density matrix formalism for neutral $K$ decays was introduced in
ref. \cite{EHNS} and reviewed in ref. \cite{emnkaons}.
The experimental value of any
observable $O$ is given by the expectation value
\be
<<O>> = Tr O\rho
\label{rhoden}
\ee

\nk where $O$ is
represented by a suitable hermitian $2 \times 2$ matrix. We
express the matrices $O_{ij}$ in the $K_{1,2}$ basis introduced
earlier. The most
commonly measured observables are the $K \rightarrow 2{\pi}$
decay rate $O_{2{\pi}}$:
\be
 O_{2\pi} =\left( \begin{array}{c} 0 \qquad 0 \\
0 \qquad 1 \end{array} \right)
\label{o2pi}
\ee

\nk and the semileptonic $K$ decay rates:
\be
  O_{\pi ^-l^+ \nu} = \left(\begin{array}{c} 1 \qquad 1 \\
1\qquad 1 \end{array}\right)
\label{semilepton}
\ee

\nk and
\be
  O_{\pi ^+l^-{\overline \nu}}  =\left( \begin{array}{c}

1 \qquad -1 \\
 -1\qquad 1 \end{array} \right)
\label{antisem}
\ee

\nk out
of which the semileptonic decay asymmetry observable $\delta$ can
be constructed
\be
\delta \equiv \frac{\Gamma (\pi^-l^+\nu) - \Gamma (\pi^+ l^-
{\overline \nu }) }{\Gamma (\pi ^- l^+ \nu ) +
\Gamma (\pi ^+ l^- {\overline \nu})}
\label{decasymm}
\ee

\nk Another variable which we discuss here for the first time in this
framework is the $K \rightarrow 3{\pi}^0$ decay rate $O_{3{\pi}}$:
\be
 O_{3\pi}
 =(0.22)
 \left( \begin{array}{c} 1 \qquad 0 \\
0 \qquad 0 \end{array} \right)
\label{o3pi}
\ee

\nk where the prefactor is determined by the measured \cite{pdg}
branching ratio for
$K_L \rightarrow 3{\pi}^0$. (Strictly speaking, there should be a
corresponding prefactor of $0.998$ in the formula (\ref{o2pi}) for
the $O_{2{\pi}}$ observable.)
\pr
It is a simple matter to combine the above formulae for $K$ decay
observables with the asymptotic solutions (24, 25) to the
non-quantum-mechanical equations of motion for the density matrix to
obtain parametrizations of the values of the observables
$R_{2\pi}^L\equiv <<O_{2\pi} >>_L$, $\delta _{L,S}$,
and $R_{3\pi}^S \equiv <<O_{3\pi}>>_S/0.22 $ :
\bea
\nn
R_{2\pi}^L &=&
|\epsilon |^2 + \frac{\gamma}{\Delta \Gamma }
+\frac{4\beta}{ |\Delta \Gamma|}
 |\epsilon| sin\phi _{\epsilon} -
 \frac{4\beta ^2 }{|\Delta \Gamma |^2} cos^2 \phi_{\epsilon}  \\
\nn
\delta_L &=&     2Re [\epsilon (1-\frac{i\beta}{Im M_{12}})] \\
\nn    \delta_S  &=&
2Re [ \epsilon (1 + \frac{i\beta}{Im M_{12}})]  \\
R_{3\pi}^S &=&
|\epsilon |^2 + \frac{\gamma} {\Delta \Gamma }
-\frac{4\beta}{|\Delta \Gamma|}
 |\epsilon| sin\phi _{\epsilon} -
 \frac{4\beta ^2 }{|\Delta \Gamma |^2} cos^2 \phi_{\epsilon}
\label{rddr}
\eea

\nk that can be compared with experiment.
\pr
We are now in a position to confront the above formalism with salient
aspects of the available data \cite{pdg}. These include the
well-measured rate for $K_L \rightarrow 2 \pi$:
\be
   \sqrt{ R_{2\pi}^L} \simeq  (2.265 \pm 0.023) \times 10^{-3}
\label{r2pi}
\ee

\nk the observed $K_L$ semileptonic decay asymmetry:
\be
 \delta _L = (3.27 \pm 0.12 ) \times 10^{-3}
\label{dell}
\ee

\nk and an upper bound on the rate for $K_S \rightarrow 3 {\pi}^0$:
\be
     \sqrt{R_{3\pi}^S} < 1.3  \times 10^{-2}
\label{r3pi}
\ee

\nk For our purposes, a more relevant quantity is the difference between
$R_{2{\pi}}^L$ and $R_{3{\pi}}^S$:
\be
\delta R \equiv R_{2\pi}-R_{3\pi}= \frac{8\beta}{|\Delta \Gamma |}
 |\epsilon| sin\phi _{\epsilon}
\label{dr}
\ee

\nk Now also available is a recent preliminary
measurement \cite{CPLEAR}
of the $K_S$
semileptonic decay asymmetry by the CPLEAR collaboration:
\be
          \delta_S \simeq (8.5 \pm
          7.6(stat) \pm 15.5(syst))\times 10^{-3}
\label{dels}
\ee

\nk Instead of ${\delta}_S$, it is more relevant to plot the difference
between $\delta_L$ and $\delta_S$:
\be
          \delta\delta \equiv
\delta _L  - \delta _S  = -\frac{8\beta} {|\Delta \Gamma |}
\frac{sin\phi _{\epsilon}}{\sqrt{1 + \tan^2 \phi _{\epsilon}}}
= -\frac{8\beta}
{|\Delta \Gamma |} sin\phi _{\epsilon}
cos\phi _{\epsilon}
\label{dd}
\ee

\pr
In
addition to these pieces of information which we include exactly in
our analysis, important measurements are also available \cite{NA31}
of the
interferences between $K_S$ and $K_L$ decays into the ${\pi}^+{\pi}^-$
and $2{\pi}^0$ final states. These are conventionally used to constrain
the phase of the $CP$-violating
mass mixing parameter $\epsilon$, which is
then compared with the value predicted on the basis of the observed
values of the $K_L - K_S$ mass and lifetime differences:
\be
\phi _{\epsilon}=arctan(\frac{2\Delta M}{\Delta \Gamma})
\label{phie}
\ee

\nk This comparison yields
\be
          {\delta}{\phi}_{\epsilon} \simeq (2.3 \pm 1.4)^0
\label{dph}
\ee

\nk which is consistent with zero and hence the absence
$CPT$ violation at the $1.5 \sigma$
level. Since we do not wish to overplay any apparent
discrepancy with $CPT$ invariance without presenting
a global analysis of the available data \cite{panz}, for
the purposes of this paper we interpret the limits
(\ref{dph}) as corresponding to $|\delta \phi _{\epsilon} |<
4.6^0 $. In our case a fit to the data is more
complicated than in conventional quantum-mechanics,
because we have three
$CPT$-violating parameters ($\alpha$, $\beta$, $\gamma$), instead of the
single quantum-mechanical parameter ${\delta}M$ that is usually
discussed. The value of ${\delta}M$ is usually obtained from a global
fit to all the available data on $K \rightarrow 2{\pi}$ decays, from
the short-time $K_S$ region through the intermediate-time interference
region to the long-time $K_L$ region. Such a complete analysis goes
beyond the scope of this paper \cite{elmn}.
However, since these data have been used to bound
the quantum-mechanical $CPT$-violating parameter
$|\delta M | \lsim 2\times 10^{-18}$ $GeV$, and since $\beta$
appears (22) in similar entries in the time evolution
matrix, we expect that such a fit would yield

\be
|\frac{\beta}{\Delta \Gamma}| \lsim ~10^{-4}~to~10^{-3}
\label{bound}
\ee

\nk depending on the values of $\alpha$ and $\gamma$.
\pr We
will present the results of our analysis in the ($\beta$, $\gamma$)
plane, since the asymptotic data that we chiefly use do not depend
on the third $CP$-violating parameter $\alpha$, which would
need to be included in a global fit. The value of the
$CP$-violating parameter $\epsilon$ is fixed at any point in the
($\beta$, $\gamma$) plane by the relatively well-determined value
(\ref{r2pi}) of $R_{2{\pi}}^L$:
\be
|\epsilon| = -\frac{2\beta}{|\Delta \Gamma |}sin\phi _{\epsilon}
+ \sqrt{\frac{4\beta ^2}{|\Delta \Gamma |^2 } - \frac{\gamma}
{|\Delta \Gamma |} + R_{2\pi}^L }
\label{eps}
\ee

\nk where we take $\phi _{\epsilon}$ from equation (39)
and
we use this formula to plot contours of $\epsilon$ in the subsequent
graphs.
The other constraints (34, 35, 36) can be expressed in
terms of $R_{2{\pi}}^L$, $\beta$ and $\gamma$:
\be
    \frac{\gamma}{|\Delta \Gamma |}  =
R_{2\pi}^L - \frac{\delta _L ^2}{4 cos^2 \phi_{\epsilon}}
- 4\delta _L tan \phi _{\epsilon} \frac{\beta}{|\Delta \Gamma |}
- 16 sin^2 \phi_{\epsilon} \frac{\beta ^2}{|\Delta \Gamma |^2}
+ 4\frac{\beta ^2}{|\Delta \Gamma |^2}
\label{ftree}
\ee
\be
    \frac{\gamma}{|\Delta \Gamma |} =    R_{2\pi}^L -
\frac{(\delta R)^2}{64 sin^2 \phi _{\epsilon}} \frac{|\Delta \Gamma|}
{\beta} - \frac{1}{2}\delta R + 4 cos^2{\phi}_{\epsilon} \frac{\beta ^2}
{|\Delta \Gamma |^2}
\label{ff}
\ee

\nk where
\be
    \delta R  =  4\delta _L tan\phi _{\epsilon}
\frac{\beta}{|\Delta \Gamma |} + 16 sin^2 \phi _{\epsilon}
\frac{\beta ^2}{|\Delta \Gamma |^2}
\label{Dr}
\ee

\nk and
\be
\delta\delta = -\frac{8\beta}{|\Delta \Gamma |} sin\phi _{\epsilon}
cos\phi _{\epsilon} = (5.2 \pm 17.3) \times 10^{-3}
\label{dd2}
\ee

\nk which we plot as bands in the ($\beta$, $\gamma$) plane.
\pr
Figures 1 and 2 show this plane on logarithmic scales for $\beta > 0$
and $ <$ $0$ respectively. We see that consistency between
$R_{2{\pi}}^L$ and the relatively well-determined value of $\delta_L$
specifies a very narrow band in the ($\beta$, $\gamma$) plane. The
origin $\beta$ = $\gamma$ = $0$, which corresponds to the conventional
state-vector analysis with $|\epsilon |$ = $2.265 \times
10^{-3}$ and without
$CPT$ violation, lies
comfortably within this band. This point can be seen
more clearly in figure 3, which shows smaller values of $\beta$ on a
linear scale. Also shown in figures 1, 2, 3 are the constraints
(44, 46), which allow relatively large values of $|\beta|$ and are also
consistent with $\beta$ = $\gamma$ = $0$ as in conventional quantum
mechanics.
\pr
The constraints (44, 46) are consistent not only
with the conventional quantum-mechanical $CP$-violating
and $CPT$-conserving solution $|\epsilon | \simeq
2.265 \times 10^{-3} $, $\beta = \gamma =0$,
but also with a purely $CP$-conserving
and $CPT$-violating solution $\epsilon =0,
\beta =-0.55 \times 10^{-3},
\gamma =  0.58 \times 10^{-5} $, as seen in fig. 4. This
radical solution
is also compatible with the indicative version
(41) of the intermediate-time constraint, though
it could well be ruled out by a more detailed
intermediate-time analysis \cite{elmn}.
As already mentioned, we do not take (40)
as significant evidence of $CPT$ violation.
\pr
\section{Speculations}
\pr
It is with some trepidation that we develop in this section the
comments made in the previous paragraph. The $CP$ violation apparently
seen in the neutral kaon system all of 28 years ago \cite{cp}
was at first a big
surprise, and did not fit naturally within the theoretical framework
then existing. However, $CPT$ violation was even more sacrosanct.
Sakharov \cite{sakh}
pointed out that $CP$ violation could solve very neatly one of
the fundamental puzzles of cosmology, and then $CP$ violation emerged
naturally from the three-generation Standard Model \cite{KM}.
So the conventional
wisdom learned to embrace $CP$ violation, and innumerable theoretical and
experimental papers have adopted it, comforted by experimental fits to
the data that did not require any
$CPT$ violation \cite{pdg}. It would take a lot to
overturn this agreeable consensus.
\pr
We are not yet equipped to do so, because we have not yet made a
complete fit to all the available experimental data, including those in
the intermediate-time region where (conventionally) the phase of the
$CP$-violating parameter $\epsilon$ is checked.
For the time being, the measurement (40) should be treated very
cautiously.
We have found
that all the available asymptotic (short- and long-time) data are
consistent
with $CP$ invariance and the generalized parametrization of $CPT$
violation that we have derived from the modification of quantum mechanics
that we have proposed\footnote{It should be remarked
that in this formalism the $T$-reversal violation
that defines the arrow of target time is of the same nature
in the microcosmos and the macrocosmos, and stems from the
unitarity of the effective low-energy world and the associated
irreversibility \cite{emnqm}
of the renormalization group flow at the
string world-sheet level \cite{zam}.}.
However, we think there
is a chink in the armour
of the conventional wisdom that needs to be explored.
As already pointed out, we cannot yet even
exclude the radical possibility that the $CP$-violating
parameter $\epsilon$ actually vanishes,
and that the effects usually ascribed
to $CP$ violation are in fact due to non-quantum-mechanical
$CPT$ violation \footnote{We are fully aware of the dramatic
consequences this outlandish possibility would have
for searches for $CP$ violation in the $B$ system, the neutron,
and atomic electric dipole moments (doomed), the axion
(demotivated) and cosmological baryogenesis (unscathed).}.
\pr
This possibility could perhaps be excluded
by refitting the available data on $K \rightarrow
2{\pi}$ using
our generalized parametrization, and/or reducing somewhat the
experimental
errors on the $CPT$-violating quantities ${\delta}{\delta}$ and
${\delta}R$. One would at least be able to refine
the present constraints on the $CPT$ violating parameters
$\beta$ and $\gamma$ and (using intermediate-time data) begin
to constrain $\alpha$.
\pr
We are
familiar with the historical fact that more
discrete symmetries are violated as one makes more
precise microscopic measurements: first $C$ and $P$
violation in the weak interactions, then $CP$ violation,
and next ...? We have argued in this paper on the basis
of string theory that $CPT$ violation could show up
at a level not far below the sensitivity of present
experimental limits. We cannot make precise, quantitative
estimates of the possible magnitude of $CPT$-violating
effects. However, they stand out as a possible
distinctive phenomenological signature of string theory.
Could the era of string phenomenology be ushered
in by such a non-perturbative, non-quantum-field-theoretical
and non-quantum-mechanical effect?

\pr
\noindent {\Large{\bf Acknowledgements}}
\pr
J.E. and N.E.M. thank L. Maiani for encouraging discussions
and for
invitations to
participate
in the {\it Second Da$\phi$ne Workshop}, Frascati
18-21 November 1992,  where preliminary
results of this work were
presented. We also thank T. Geralis and M. Fidecaro
for useful communications about CPLEAR results.
The work of D.V.N. is partially supported by DOE grant
DE-FG05-91-ER-40633 and by a grant from Conoco Inc.

\newpage
{\Large {\bf Figure Captions}}
\pr

\nk {\Large {\bf Figure 1 }}- The
$(\beta, \gamma)$ plane on
a logarithmic
scale for $\beta > 0$. We plot
contours of the conventional
$CP$-violating parameter $|\epsilon |$, evaluated from the
$K_L \rightarrow 2\pi $ decay rate. The dashed-double-dotted
band is that allowed
at the one-standard-deviation level by the comparison
between measurements of the $K_L \rightarrow 2\pi $ decay rate
and the $K_L$ semileptonic decay asymmetry $\delta _L $.
The dashed line delineates the boundary
of the region allowed by the
present experimental upper limit on $K_S \rightarrow 3\pi ^0$
decays $(R_{2\pi}^L-R_{3\pi}^S)$
and a solid line delineates the boundary
of the region allowed by
a recent preliminary measurement of the $K_S$
semileptonic decay asymmetry $\delta _S$.
A wavy line bounds approximately the region of $|\beta |$
which may be prohibited by intermediate-time measurements of
$K \rightarrow 2\pi$ decays.
\pr

\nk {\Large {\bf Figure 2 }}-  As in Fig. 1, on a
logarithmic scale for $\beta < 0$.

\pr
\nk {\Large {\bf Figure 3 }}-  As in Fig. 1, on a
linear scale for the neighborhood of $\beta=0$.
\pr

\nk {\Large {\bf Figure 4 }}-  As in Fig. 1, in a
blown-up region around $\beta =-0.55 \times 10^{-3}$, $\gamma
= 0.58 \times 10^{-5}$, corresponding to the absence
of $CP$ violation : $\epsilon = 0$.

\end{document}